\documentstyle[12pt]{article}
\begin{document}

\title{Linear optical implementation of a single mode quantum filter and
generation of multi-photon polarization entangled state}
\author{XuBo Zou, K. Pahlke and W. Mathis  \\
Electromagnetic Theory Group at THT\\
 Department of Electrical
Engineering University of Hannover, Germany}
\date{}

\maketitle

\begin{abstract}
{\normalsize We propose a scheme to implement a single-mode
quantum filter, which selectively eliminates the one-photon state
in a quantum state
$\alpha|0\rangle+\beta|1\rangle+\gamma|2\rangle$. The vacuum state
and the two photon state are transmitted without any change. This
scheme requires single-photon sources, linear optical elements and
photon detectors. Furthermore we demonstrate, how this filter can
be used to realize a two-qubit projective measurement and to
generate multi-photon polarization entangled states.

 }
\end{abstract}
The generation of polarization entangled quantum states of
individual photons is an important challenge in the field of
information processing with quantum optical systems. In particular
the robust nature of this kind of quantum states in respect of
decoherence effects provides an advantage over other quantum
systems. Related experiments opened a wide field of research
\cite{daa}. Two-photon polarization entangled states can be used
as a simple resource of entanglement. An experimental realization
of these states was demonstrated with the parametric
down-conversion technique in a nonlinear optical crystal
\cite{kwait}. These two-photon polarization entangled states can
be used to test quantum nonlocality \cite{daa} and to implement
quantum information protocols like quantum teleportation
 \cite{db}, quantum dense coding \cite{km} and quantum cryptography
 \cite{ds}. The entanglement of three-photon states and four-photon states
was realized experimentally \cite{dj}. These states can be used to
demonstrate the extreme contradiction between local realism and
quantum mechanics \cite{dj1} in the so-called
Greenberger-Horne-Zeilinger-state (GHZ-state)
\cite{ghz}.\\
So far polarization entangled photon states have only been
produced randomly, since there is no way to demonstrate their
generation without a measurement and destruction of the outgoing
quantum state \cite{kb}. This is a severe obstacle for further
applications of this state to be used as the input state for
following experiments. Some quantum protocols like error
correction were designed for maximally entangled quantum states
without random entanglement \cite{dbo}. Thus, there is a strong
need in entangled photon sources, which generate maximally
polarization entangled multi-photon states in a controllable way.
Recently, significant progress was achieved by a proposal to
implement probabilistic quantum logic gates by the use of linear
optical elements and photon detectors \cite{klm}. Although the
scheme in Ref.\cite{klm} contains only linear optical elements,
the optical network is complex and major stability and mode
matching problems can be expected in their construction. Several
less complicated schemes were presented
 \cite{ra,ru,zou} to implement the quantum logic operation with a
slightly lower probability of success. But, all these schemes can
be used directly to generate two-photon polarization entangled
states from single-photon sources with a probability, which is not
greater than $1/16$. Another simple scheme \cite{zou1} was
proposed to generate two-photon polarization entangled states and
entangled $N$-photon states from single-photon sources. Recently,
a single-photon quantum nondemolition device was proposed for
signaling the presence of a single photon in a particular input
state $\alpha|0\rangle+\beta|1\rangle+\gamma|2\rangle$ without
destroying it \cite{yy1}. In Ref.\cite{hh} Hofmann et al proposed
a scheme to realize a two-qubit projective measurement
\begin{eqnarray}
P_e=|H,H\rangle\langle H,H|+|V,V\rangle\langle V,V| \,.\label{1}
\end{eqnarray}
The two-mode quantum states
$|H,H\rangle=|H\rangle_1|H\rangle_2=|H\rangle_1\otimes|H\rangle_2$
and
$|V,V\rangle=|V\rangle_1|V\rangle_2=|V\rangle_1\otimes|V\rangle_2$
are constructed from single-mode quantum states by the tensor
product $\otimes$. The quantum state of a horizontally
(vertically) polarized photon in the mode $i$ is indicated by
$|H\rangle_i$($|V\rangle_i$). This measurement (\ref{1}) projects
the input state of two polarized photons into the two-dimensional
subspace of equal polarization (both horizontal or both vertical).
These proposals \cite{klm,ra,ru,zou,zou1} show that photon number
detection can be used to generate polarization entanglement, which
is in general associated with nonlinear effects. In the present
paper, we propose a scheme to realize the two-qubit projective
measurement (\ref{1}), which gives compared to the scheme
\cite{hh}
a four times greater probability of the outcome.\\
This paper is organized as follows. At first we propose a scheme
to implement the single-mode quantum filter, which eliminates the
one-photon state in a particular input state
$\alpha|0\rangle+\beta|1\rangle+\gamma|2\rangle$. Then this filter
is applied to realize a two-qubit projective measurement (\ref{1})
and to show, how multi-photon polarization entangled states can be
efficiently generated step by step.\\

In contrast to the scheme \cite{hh} our single-mode quantum filter
was inspired by quantum teleportation protocols \cite{bb}, which
make our scheme more efficient for the implementation of the
projective measurement (\ref{1}). Figure 1 shows the scheme of
this single-mode quantum filter, which eliminates the one-photon
state and transmits the vacuum state and the two-photon state of
the input state
\begin{eqnarray}
\Psi_{in}=\alpha|0\rangle_1+\beta|1\rangle_1+\gamma|2\rangle_1\,.\label{2}
\end{eqnarray}
A similar setup was used to implement the conditional quantum
logic gate \cite{zou}. But here, a symmetric beam splitter $BS_1$
is required. By emitting two single photons
$|\Psi_{23}\rangle=|1\rangle_2|1\rangle_3$ into this symmetric
beam splitter $BS_1$ an entangled photon channel is generated:
\begin{eqnarray}
\Psi^{\prime}_{23}=
\frac{1}{\sqrt{2}}(|2\rangle_{2^{\prime}}|0\rangle_{3^{\prime}}-|0\rangle_{2^{\prime}}|2\rangle_{3^{\prime}})\,.
\label{3} \end{eqnarray}
Our idea is to use the field state
$|\Psi_{in}\rangle|\Psi^{\prime}_{23}\rangle$ directly for the
quantum filter operation. The output mode $2^{\prime}$ of the beam
splitter $BS_1$ is one of the input modes of the second symmetric
beam splitter $BS_2$. The mode $1$, which is chosen to be filtered
out, is the
second input mode of the beam splitter $BS_2$. Thus, the quantum
state transforms to
\begin{eqnarray}
\Psi_{o}&=&|1\rangle_{1^{\prime\prime}}
|1\rangle_{2^{\prime\prime}}( \alpha|0\rangle_{3^{\prime}}+
\gamma|2\rangle_{3^{\prime}})/2 +|0\rangle_{1^{\prime\prime}}
|2\rangle_{2^{\prime\prime}}( \alpha|0\rangle_{3^{\prime}}-
\gamma|2\rangle_{3^{\prime}})/2\sqrt{2}\nonumber\\
&&+|2\rangle_{1^{\prime\prime}} |0\rangle_{2^{\prime\prime}}(
\alpha|0\rangle_{3^{\prime}}-
\gamma|2\rangle_{3^{\prime}})/2\sqrt{2} + \Psi_{other}\,.\label{4}
\end{eqnarray} The other terms $\Psi_{other}$ of the total
quantum state $\Psi_{o}$ do not contribute to the three events,
which we consider by a photon number measurement on the modes
$1^{\prime\prime}$ and $2^{\prime\prime}$ by the detectors $D_1$
and $D_2$. If each detector detects one photon, the state is
projected into the (unnormalized) state
\begin{eqnarray}
\Psi_{out}=\alpha|0\rangle+\gamma|2\rangle\,. \label{5}
\end{eqnarray} If $D_1$ detects two photons and $D_2$ does not
detect any photon or vice versa, the state is projected into the
(unnormalized) state
\begin{eqnarray}
\Psi_{out}^{\prime}=\alpha|0\rangle-\gamma|2\rangle \,.\label{6}
\end{eqnarray} In order to generate the quantum state (\ref{5}), a
$\pi/2$-phase shifter is needed to change the sign of the state
$|2\rangle$. Then the desired transformation
\begin{eqnarray}
\alpha|0\rangle+\beta|1\rangle+\gamma|2\rangle\longrightarrow\alpha|0\rangle+\gamma|2\rangle
\label{7} \end{eqnarray} will be obtained. This transformation
demonstrates that the proposed setup, which is shown in Fig.1,
definitely implements the quantum state filter. The probability of
this outcome is $\frac{|\alpha|^2+|\gamma|^2}{2}$.\\
In Ref.\cite{yy1} Kok et al proposed a different quantum
nondemolition device, which transmits only the single-photon
state. Thus, these schemes are within the constraint of a
particular input state in that respect complementary. Quantum
state filters of that kind may find a wide range of application,
from quantum projective measurements to the generation of multi
photon quantum states.\\

Now we use our quantum filter concept to implement the two-qubit
projective measurement (\ref{1}). We will demonstrate, that the
input state
\begin{eqnarray}
\Phi_{in}=c_0|H\rangle_1|H\rangle_2+c_1|H\rangle_1|V\rangle_2+c_2|V\rangle_1|H
\rangle_2+c_3|V\rangle_1|V\rangle_2 \label{8}
\end{eqnarray} will be transformed into the (unnormalized) state
\begin{eqnarray}
\Phi_{out}=c_0|H\rangle_1|H\rangle_2+c_3|V\rangle_1|V\rangle_2\,.
\label{9}
\end{eqnarray}
Figure 2 shows the required experimental setup to realize this
projective measurement by two quantum filters, like they are shown
in Fig.1. The half-wave plate $HWP_{90}$ rotates the polarization
of the mode $1$ by $\pi/2$. The state (\ref{8}) becomes
\begin{eqnarray}
\Phi_{1}=c_0|V\rangle_1|H\rangle_2+c_1|V\rangle_1|V\rangle_2+c_2|H\rangle_1|
H\rangle_2+c_3|H\rangle_1|V\rangle_2\,. \label{10} \end{eqnarray}
Then mode $1$ and the mode $2$ are forwarded to a polarization
beam splitter $PBS_1$ to transmit the $H$-polarized photon and to
reflect the $V$-polarized photon. The system state evolves into
\begin{eqnarray}
\Phi_{2}=c_0|HV\rangle_{2^{\prime}}+c_1|V\rangle_{1^{\prime}}|V\rangle_{2^{\prime}}
+c_2|H\rangle_{1^{\prime}}|H\rangle_{2^{\prime}}
+c_3|HV\rangle_{1^{\prime}}\,.\label{11} \end{eqnarray} The
quantum state representation of two photons in the same mode $i$
is defined as:
$|HV\rangle_i=|VH\rangle_i=|H\rangle_i\otimes|V\rangle_i$. Another
one half-wave plate ($HWP_{45}$) rotates the polarization of the
mode $1^{\prime}$ by $\pi/4$. This corresponds to the
transformations
$|H\rangle\longrightarrow\frac{1}{\sqrt{2}}(|H\rangle+|V\rangle)$
and
$|V\rangle\longrightarrow\frac{1}{\sqrt{2}}(|H\rangle-|V\rangle)$.
The quantum state of the system is transformed into
\begin{eqnarray}
\Phi_{3}&=&c_0|HV\rangle_{2^{\prime}}+\frac{1}{\sqrt{2}}|H\rangle_{3}(c_1|V
\rangle_{2^{\prime}}+c_2|H\rangle_{2^{\prime}})
\nonumber\\
&&+\frac{1}{\sqrt{2}}|V\rangle_{3}(-c_1|V\rangle_{2^{\prime}}+c_2|H\rangle_{2^{\prime}})
+\frac{c_3}{\sqrt{2}}(|2H\rangle_{3}-|2V\rangle_{3})\,. \label{12}
\end{eqnarray} The mode $3$ is injected into another polarization
beam splitter $PBS_2$ so that the system state becomes
\begin{eqnarray}
\Phi_{4}&=&c_0|HV\rangle_{2^{\prime}}+\frac{1}{\sqrt{2}}|H\rangle_{4}(c_1|V
\rangle_{2^{\prime}}+c_2|H\rangle_{2^{\prime}})
\nonumber\\
&&+\frac{1}{\sqrt{2}}|V\rangle_{5}(-c_1|V\rangle_{2^{\prime}}+c_2|H\rangle_{2^{\prime}})
+\frac{c_3}{\sqrt{2}}(|2H\rangle_{4}-|2V\rangle_{5})\,. \label{13}
\end{eqnarray} The mode $4$ and the mode $5$ pass a single-mode quantum
filter, like it is shown in Fig.1, to eliminate the one-photon
state of the mode $4$ and the mode $5$. This filtering process
transforms the state (\ref{13}) to the state
\begin{eqnarray}
\Phi_{5}=c_0|HV\rangle_{2^{\prime}}
+\frac{c_3}{\sqrt{2}}(|2H\rangle_{4^{\prime}}-|2V\rangle_{5^{\prime}})\,.
\label{14} \end{eqnarray} The mode $4^{\prime}$ and the mode
$5^{\prime}$ are two input modes of the polarization beam splitter
$PBS_3$. That's why the system state transforms to
\begin{eqnarray}
\Phi_{6}=c_0|HV\rangle_{2^{\prime}}
+\frac{c_3}{\sqrt{2}}(|2H\rangle_{3^{\prime}}-|2V\rangle_{3^{\prime}})\,.
\label{15} \end{eqnarray} The polarization of the mode
$3^{\prime}$ is rotated by $\pi/4$ to get
\begin{eqnarray}
\Phi_{7}=c_0|HV\rangle_{2^{\prime}} +c_3(|HV\rangle_{6})\,.
\label{16}
\end{eqnarray} And finally the mode $6$ and the mode $2^{\prime}$ pass
through a polarization beam splitter $PBS_4$ in order to obtain
the output state (\ref{9}) of the implemented two-qubit projective
measurement (\ref{1}). The probability of this outcome is
$(|c_0|^2+|c_3|^2)/4$, which
is four times greater than what the proposal \cite{hh} makes possible.\\

Now we show how the multi-photon polarization entangled states can
be efficiently generated. At first we consider the generation of
the two-photon polarization entangled state
$(|H\rangle|H\rangle+|V\rangle|V\rangle)/\sqrt{2}$. We assume that
the two polarized photons form the initial state
$(|H\rangle_1+|V\rangle_1)/\sqrt{2}\otimes(|H\rangle_2+|V\rangle_2)/\sqrt{2}$.
These two photons are injected into two input ports of the setup,
which is shown in Fig.2. Notice that this setup projects the input
state into the two-dimensional subspace of the identical
horizontal or vertical polarization. After passing through the
setup, the initial state is projected into the two-photon
polarization entangled state
$(|H\rangle|H\rangle+|V\rangle|V\rangle)/\sqrt{2}$. The pobability
of this outcome is $1/8$. In the following, we consider the
generation of a multi-photon polarization GHZ-state. We assume
that a $N-1$-photon polarization GHZ-state
$(|H\rangle_1\cdots|H\rangle_{N-1}+|V\rangle_1\cdots|V\rangle_{N-1})/\sqrt{2}$
and the $N$th independent polarization photon are prepared
initially in the state $(|H\rangle_N+|V\rangle_N)/\sqrt{2}$. If
the $(N-1)$th photon and the $N$th polarization photon inject into
two input ports of the setup, which is shown in Fig.2., the
$N$-photon polarization GHZ-state
$(|H\rangle_1\cdots|H\rangle_{N}+|V\rangle_1\cdots|V\rangle_{N})/\sqrt{2}$
will be obtained. The probability of this outcome is $1/8$. This
demonstrates that $N$-photon polarization entangled GHZ-states can
be generated from single-photon sources step by step. The
probability of the outcome
of the total process is $\frac{1}{8^{N-1}}$.\\

In summary, we have proposed a scheme to implement a quantum
filter, which eliminates the one-photon state and transmits the
vacuum state and the two-photon state in the particular input
state $\alpha|0\rangle+\beta|1\rangle+\gamma|2\rangle$. We
demonstrated that the quantum filter can be applied to implement a
two-qubit projective measurement. This two-qubit projective
measurement was also proposed by Hofmann et al \cite{hh}, but our
scheme provides a four times bigger probability of outcome.\\
In the present scheme, the multi-photon polarization entanglement
is produced in a controllable way, so that it can be used to
generate multi-photon polarization entanglement for some quantum
protocols, which are designed for maximally entangled quantum
states without random entanglement \cite{dbo}. In order to compare
the efficiency of our scheme to generate the $N$-photon
polarization GHZ-state with the scheme of Knill et al \cite{klm}
the same requirements on the sources should be posed. In the case
of single-photon sources the scheme of Knill et al gives this
outcome with the probability $\frac{1}{16^{N-1}}$. Indeed, in the
case of an entangled photon source the scheme \cite{klm} becomes
nearly deterministic, but
the main problem is then shifted to the sources.\\
One of the difficulties of our scheme in respect to an
experimental demonstration consists in the requirement on the
sensitivity of the detectors. These detectors should be capable to
distinguish between one-photon events and two-photon events.
Recently, the development of experimental techniques in this field
made considerable progress. A photon detector based on visible
light photon counter was reported, which can distinguish between a
single-photon incidence and a two-photon incidence with a high
quantum efficiency, a good time resolution and a low bit-error
rate \cite{yyy}. Another difficulty is the availability of
single-photon sources. Several triggered single-photon sources are
available, which operate by means of fluorescence from a single
molecule \cite{cbb} or a single quantum dot \cite{cs,pm}. They
exhibit a very good performance. More recently a deterministic
single-photon source for distributed quantum networks was reported
\cite{rempe}, which can emit a sequence of single photons on
demand from a single three-level atom strongly coupled to a
high-finesse optical cavity. However, our scheme needs a
synchronized arrival of many single photons into input ports of
many beam splitters. An experimental realization will be
challenging.\\

\begin{flushleft}

{\Large \bf Figure Captions}

\vspace{\baselineskip}

{\bf Figure 1.} The schematic shows the implementation of the
single-mode quantum filter. $BS_i$(i=1,2) denotes the symmetric
beam splitters and $D_i$(i=1,2,3) are photon number detectors. The
$\pi/2$ phase shifter is denoted by $P$.

{\bf Figure 2.} This scheme shows the implemention of the
projective measurement (\ref{1}). The half-wave plates $HWP_{45}$
and $HWP_{90}$ rotate the horizontal and vertical polarization by
$\pi/4$ and $\pi/2$. The polarization beam splitters
($PBS_i$(i=1,2,3,4)) transmit H-photons and reflect V-Photons.
$F_i$ (i=1,2) denotes the single mode quantum filter, like it is
shown in Fig.1.

\end{flushleft}


\begin{thebibliography}{99}
\bibitem{daa}D. Bouwmeester, A. Ekert and A. Zeilinger, The physics of Quantum
Information( Springer-Verlag, Berlin, 2000)
\bibitem{kwait}P. G. Kwiat, K. Mattle, H. Weinfurter, A. Zeilinger, A. V. Sergienko,
and Y. H. Shih, Phys. Rev. Lett. 75, 4337 (1995).
\bibitem{db}D. Bouwmeester et al , Nature(London) {\bf 390} 575 (1997)
\bibitem{km}K. Mattle et al,  Phys. Rev.
Lett. {\bf 76}, 4656 (1996).
\bibitem{ds} D. S. Naik et al  Phys. Rev. Lett.  {\bf 84},
4733 (2000); W. Tittel et al  Phys. Rev. Lett.  {\bf 84}, 4737
(2000).
\bibitem{dj} D. Bouwmeester et al., Phys. Rev. Lett.  {\bf 82},
1345 (1999); J. W. Pan et al, Phys. Rev. Lett.  {\bf 86}, 4435
(2001).
\bibitem{dj1}J.-W. Pan et al., Nature (London) 403, 515 (2000).
\bibitem{ghz}
D. M. Greenberger, M. A. Horne, and A. Zeilinger, in Bell's
Theorem, Quantum Theory, and Conceptions of the Universe, edited
by M. Kafatos (Kluwer Academics, Dordrecht, The Netherlands 1989),
pp. 73–76; D. M. Greenberger, M. A. Horne, A. Shimony, and A.
Zeilinger, Am. J. Phys. 58, 1131 (1990)
\bibitem{kb}P. Kok and S. Braunstein, quant-ph/0001080
\bibitem{dbo}D. Bouwmeester, quant-ph/0006108
\bibitem{klm}E. Knill, R. Laflamme and G. Milburn, Nature(London) {\bf
409}, 46 (2001).
\bibitem{ra}T.C.Ralph, A.G.White, W.J.Munro, G.J.Milburn, quant-ph/0108049
\bibitem{ru}Terry Rudolph, Jian-Wei Pan, quant-ph/0108056
\bibitem{zou}XuBo Zou, K. Pahlke and W. Mathis, Phys. Rev. A 65, 064305 (2002)
\bibitem{zou1}Hwang Lee, Pieter Kok, Nicolas J. Cerf, Jonathan P.
Dowling,Phys. Rev. A 65,030101 (2002); Jaromir Fiurasek, Phys.
Rev. A 65, 053818 (2002); XuBo Zou, K. Pahlke and W. Mathis, Phys.
Rev. A 66, 014102 (2002)
\bibitem{yy1}Pieter Kok, Hwang Lee, Jonathan P. Dowling
quant-ph/0202046.
\bibitem{hh}
Holger F. Hofmann and Shigeki Takeuchi, Phys. Rev. Lett. 88, 147901 (2002)
\bibitem{bb}C.H. Bennett, G. Brassard, C. Crépeau, R. Jozsa, A. Peres, and W. Wootters,
Phys. Rev. Lett. 70, 1895 (1993).
\bibitem{yyy}Jungsang Kim, Shigeki Takeuchi, Yoshihisa Yamamoto, and Henry H.
Hogue, Applied Physics Letters, 74, 902 (1999)
\bibitem{cbb}Christian Brunel, Brahim Lounis, Philippe Tamarat, and Michel Orrit
,Phys. Rev. Lett. 83, 2722 (1999).
\bibitem{cs}Charles Santori et al., Phys. Rev. Lett. 86, 1502 (2001)
\bibitem{pm}P. Michler et al, Science 290, 2282(2000)
\bibitem{rempe} Axel Kuhn, Markus Hennrich, Gerhard Rempe,
quant-ph/0204147.
\end{thebibliography}
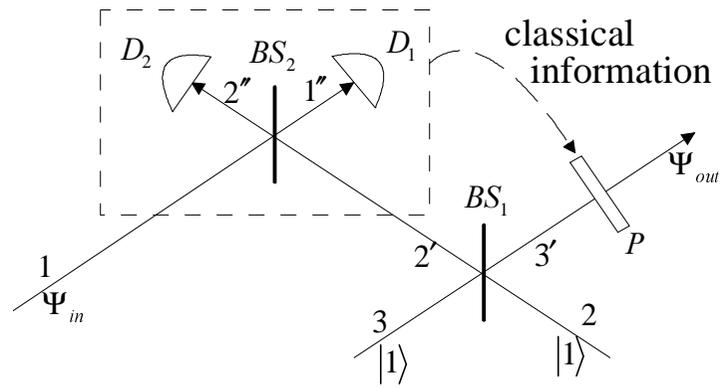
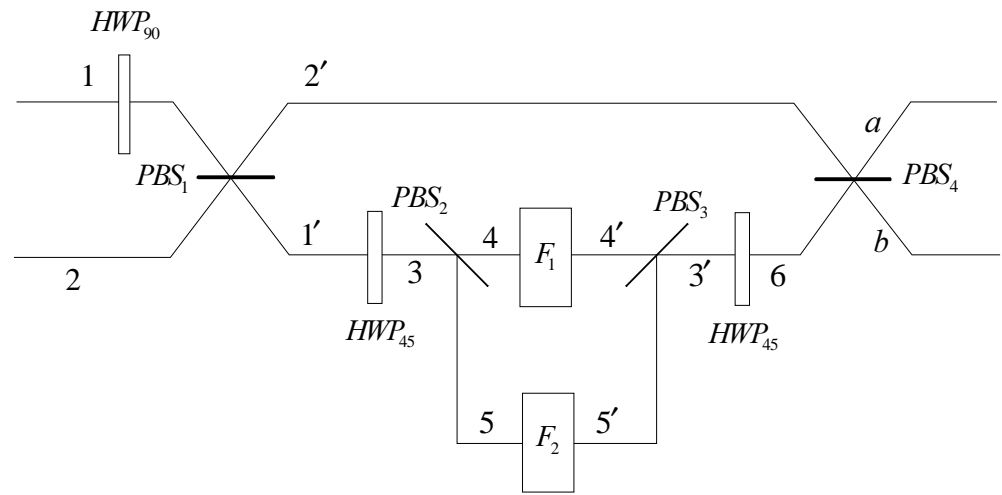
\end{document}